\documentclass[12pt]{article}
\usepackage{amsmath}
\usepackage{color}
\usepackage{graphicx}
\begin{document}
\baselineskip=18 pt
\begin{center}
{\large{\bf Investigation of spin-$0$ massive charged particle subject to a homogeneous magnetic field with potentials in a topologically trivial flat class of G\"{o}del-type space-time }}
\end{center}

\vspace{.5cm}

\begin{center}
{\bf Faizuddin Ahmed}\footnote{faizuddinahmed15@gmail.com ; faiz4U.enter@rediffmail.com}\\
{\bf Ajmal College of Arts and Science, Dhubri-783324, Assam, India}
\end{center}

\vspace{.5cm}

\begin{abstract}

In this paper, we investigate relativistic quantum dynamics of spin-$0$ massive charged particle subject to a homogeneous magnetic field in the G\"{o}del-type space-time with potentials. We solve the Klein-Gordon equation subject to a homogeneous magnetic field in a topologically trivial flat class of G\"{o}del-type space-time in the presence of a Cornell-type scalar and Coulomb-type vector potentials and analyze the effects on the energy eigenvalues and eigenfunctions.

\end{abstract}

\vspace{.3cm}


\section{Introduction}

The first solution to the Einstein's field equations containing closed time-like curves is the cylindrical symmetry G\"{o}del rotating Universe \cite{KG}. Reboucas {\it et al.} \cite{MJR,MJR2,MOG} investigated the G\"{o}del-type solutions characterized by vorticity, which represents a generalization of the original G\"{o}del metric with possible sources and analyzed the problem of causality. The line element of G\"{o}del-type solution is given by
\begin{equation}
ds^2=-\left (dt+A_{i}\,dx^{i} \right)^2+\delta_{ij}\,dx^{i}\,dx^{j},
\label{godel}
\end{equation}
where the spatial coordinates of the space-time are represented by $x^{i}$ and $i,j=1,2,3$. The different classes of G\"{o}del-type solutions have discussed in \cite{EPJC4}.

Investigation of relativistic quantum dynamics of spin-zero and spin-half particles in G\"{o}del Universe and G\"{o}del-type space-times as well the Schwarzschild, the Kerr black holes solution has been addressed by several authors. The study of relativistic wave-equations, particularly, the Klein-Gordon and Dirac equations in the background of G\"{o}del-type space-time was first studied in \cite{BDF}. The close relation between the relativistic energy levels of a scalar particle in the Som-Raychaudhuri space-time with the Landau levels was studied in \cite{ND}. The same problem in the Som-Raychaudhuri space-time was investigated and compared with the Landau levels \cite{SD}. The Klein-Gordon equation in the background of G\"{o}del-type space-times with a cosmic string was studied in \cite{JC}, and analyzed the similarity of the energy levels with Landau levels in flat space. The Klein-Gordon oscillator in the background of G\"{o}del-type space-time under the influence of topological defects was studied in \cite{JC2}. The relativistic quantum dynamics of scalar particle in the presence of an external fields in the Som-Raychaudhuri space-time under the influence of topological defects was studied in \cite{ZW}. The relativistic quantum motion of spin-$0$ particles in a flat class of G\"{o}del-type space-time was studied in \cite{EPJC2}. The study of spin-$0$ system of DKP equation in a flat class of G\"{o}del-type space-time was studied in \cite{MPLA}. In all the above system, the influence of topological defects and vorticity parameter characterizing the space-time on the relativistic energy eigenvalues was analyzed. Linear confinement of a scalar particle in the Som-Raychaudhuri space-time with cosmic string was studied in \cite{RR1} (see also, \cite{Ahmed1}). The behavior of scalar particle with Yukawa-like confining potential in the Som-Raychaudhuri space-time in the presence of topological defects was investigated in \cite{ME}. Ground state of a bosonic massive charged particle in the presence of an external fields in a G\"{o}del-type space-time was investigated in \cite{EOS} (see also, \cite{Ahmed2}). The relativistic quantum dynamics of spin-zero particles in 4D curved space-time with the cosmic string subject to a homogeneous magnetic field was studied in \cite{Ahmed3}. In addition, the relativistic wave-equations in (1+2)-dimensional rotational symmetry space-time background was investigated in \cite{AOP,AOP2,GERG,MPLA2,AOP3,GERG2}.

Furthermore, Dirac and Weyl fermions in the background of the Som-Raychaudhuri space-times in the presence of topological defects with torsion was studied in \cite{GQG}. Weyl fermions in the background of the Som-Raychaudhuri space-times in the presence of topological defects was studied in \cite{GQG2} (see, Refs. \cite{SGF,AH}). The relativistic wave-equations for spin-half particles in the Melvin space-time, a space-time where the metric is determined by a magnetic field was studied in \cite{LCNS}. The fermi field and Dirac oscillator in the Som-Raychaudhuri space-time was studied in \cite{MM}. The Fermi field with scalar and vector potentials in the Som-Raychaudhuri space-time was investigated in \cite{PS}. The Dirac particles in a flat class of G\"{o}del-type space-time was studied in \cite{EPJC4}. Dirac fermions in (1+2)-dimensional rotational symmetry space-time background was investigated in \cite{Ahmed4}.

The relativistic quantum dynamics of a scalar particle subject to different confining potentials have been studied in several areas of physics by various authors. The relativistic quantum dynamics of scalar particles subject to Coulomb-type potential was investigated in \cite{QWC,HM,FY,ALC}. It is worth mentioning studies that have dealt with Coulomb-type potential in the propagation of gravitational waves \cite{HA}, quark models \cite{CLC}, and relativistic quantum mechanics \cite{HWC,VRK,ERFM,KB}. Linear confinement of scalar particles in a flat class of G\"{o}del-type space-time, were studied in \cite{EPJC3}. The Klein-Gordon equation with vector and scalar potentials of Coulomb-type under the influence of non-inertial effect in cosmic string space-time was studied in \cite{LCNS2}. The Klein-Gordon oscillator in the presence of Coulomb-type potential in the background space-time generated by a cosmic string was studied in \cite{KB,KB2}. Other works on the relativistic quantum dynamics are the Klein-Gordon scalar field subject to a Cornell-type potential \cite{MH}, and survey on the Klein-Gordon equation in a G\"{o}del-type space-time \cite{MS}.

Our aim in this paper is to investigate the quantum effects on bosonic massive charged particle by solving the Klein-Gordon equation subject to a homogeneous magnetic field in the presence of a Cornell-type scalar and Coulomb-type vector potentials in G\"{o}del-type space-time. We see that the presence of magnetic field as well as various potential modifies the energy spectrum.

\section{\bf Bosonic charged particle : The KG-Equation}

The relativistic quantum dynamics of a charged particle of modifying mass $ m \rightarrow m+S$, where $S$ is the scalar potential is described by the following equation \cite{ERFM}
\begin{equation}
\left [\frac{1}{\sqrt{-g}}\,D_{\mu} (\sqrt{-g}\,g^{\mu\nu}\,D_{\nu})-(m+S)^2-\xi\,R \right]\,\Psi=0,
\label{1}
\end{equation}
where $g$ is the determinant of metric tensor with $g^{\mu\nu}$ its inverse, $D_{\mu}=\partial_{\mu}-i\,e\,A_{\mu}$ is the minimal substitution, $e$ is the electric charge and $A_{\mu}$ is the electromagnetic four-vector potential, and $\xi$ is the non-minimal coupling constant with the background curvature.

We choose the electromagnetic four-vector potential $A_{\mu}=(-V,\vec{A})$ with \cite{GBA}
\begin{equation}
A_{y}=-x\,B_0\quad ,\quad \vec{A}=(0,A_y,0)
\label{2}
\end{equation}
such that the constant magnetic field is along the axis $\vec{B}=\vec{\nabla}\times \vec{A}=-B_0\,\hat{z}$.

Consider the following stationary space-time \cite{CTP} (see, \cite{EPJC4,EPJC2,MPLA,EPJC3,CTP2}) in the Cartesian coordinates $(x^0=t, x^1=x, x^2=y, x^3=z)$ is given by
\begin{equation}
ds^2=-\left (dt+\alpha_0\,x\,dy \right)^2+\delta_{ij}\,dx^i\,dx^j,
\label{3}
\end{equation}
where $\alpha_0 > 0$ is a real positive constant. In Ref. \cite{EPJC4}, we have discussed different classes of G\"{o}del-type space-time. For the space-time geometry (\ref{3}), it belongs to a linear or flat class of G\"{o}del-type metrics. The parameter $\alpha_0=2\,\Omega$ where, $\Omega$ characterize the vorticity parameter of the space-time. For $\Omega \rightarrow 0$, the study space-time reduces to four-dimensional flat Minkowski metric.

The determinant of the corresponding metric tensor $g_{\mu\nu}$ is 
\begin{equation}
det\;g=-1.
\label{4}
\end{equation}
The scalar curvature of the metric is 
\begin{equation}
R=\frac{\alpha_{0}^2}{2}=2\,\Omega^2.
\label{5}
\end{equation}

For the space-time geometry (\ref{3}), the equation (\ref{1}) becomes
\begin{eqnarray}
&&[-(\frac{\partial}{\partial t}+i\,e\,V)^2+\left\{(\frac{\partial}{\partial y}-i\,e\,A_y)-2\,\Omega\,x\,(\frac{\partial}{\partial t}+i\,e\,V)\right \}^2+\frac{\partial^2}{\partial x^2}+\frac{\partial^2}{\partial z^2}\nonumber\\
&&-(m+S)^2-2\,\xi\,\Omega^2]\,\Psi=0.
\label{6}
\end{eqnarray}
Since the metric is independent of $t, y, z$. One can choose the following ansatz for the function $\Psi$
\begin{equation}
\Psi (t, x, y, z)=e^{i\,(-E\,t+l\,y+k\,z)}\,\psi(x),
\label{7}
\end{equation}
where $E$ is the total energy of the particle, $l=0,\pm\,1,\pm\,2,...$ are the eigenvalues of the $y$-component operator, and $-\infty < k < \infty $ is the eigenvalues of the $z$-component operator. 

Substituting the ansatz Eq. (\ref{7}) into the Eq. (\ref{6}), we obtain the following differential equation for $\psi (x)$ :
\begin{eqnarray}
&&[\frac{d^2}{dx^2}+(E-e\,V)^2-k^2-\{(l+e\,B_0\,x)+2\,\Omega\,x\,(E-e\,V)\}^2\nonumber\\
&&-(m+S)^2-2\,\xi\,\Omega^2]\,\psi(x)=0.
\label{8}
\end{eqnarray}

\subsection{Interaction with Cornell-type and Coulomb-type potentials} 

Here we study a spin-$0$ massive charged particle by solving the Klein-Gordon equation in the presence of an external fields in a flat class of G\"{o}del-type space-time subject to a Cornell-type scalar and Coulomb-type vector potentials. We obtain the energy eigenvalues and eigenfunctions and analyze the effects due to various physical parameters.

The Cornell-type potential contains a confining (linear) term besides the Coulomb interaction and has been successfully accounted for the particle physics data \cite{DHP}. This type of potential is a particular case of the quark-antiquark interaction, which has one more harmonic-type term \cite{MKB}. The Coulomb potential is responsible by the interaction at small distances and the linear potential leads to the confinement. The quark-antiquark interaction potential has been studied in the ground state of three quarks \cite{CA}, and systems of bound heavy quarks \cite{JDG,EE,EE2}. This type of interaction has been studied by several authors (\cite{ZW,MPLA2,ERFM,MSC,RLLV6,RLLV2,RLLV3,RLLV4,RLLV5,LFD}).

We consider the scalar $S$ to be Cornell-type \cite{ERFM}
\begin{equation}
S=\frac{\eta_c}{x}+\eta_{L}\,x,
\label{9}
\end{equation}
where $\eta_{c},\eta_{L}$ are the Coulombic and confining potential constants, respectively.

Another potential that we are interest here is the Coulomb-type potential which we discussed in the introduction. Therefore, the Coulomb-type vector potential is given by
\begin{equation}
V=\frac{\xi_c}{x},
\label{vector}
\end{equation}
where $\xi_{c}$ is the Coulombic potential constants.

Substituting the potentials (\ref{9}) and (\ref{vector}) into the Eq. (\ref{8}), we obatin the following equation:
\begin{equation}
\left [\frac{d^2}{dx^2} + \lambda-\omega^2\,x^2-\frac{j^2}{x^2}-\frac{a}{x}-b\,x \right]\,\psi(x)=0,
\label{10}
\end{equation}
where we have defined
\begin{eqnarray}
&&\lambda=E^2-m^2-k^2-2\,\eta_c\,\eta_L-2\,\xi\,\Omega^2-(l-2\,e\,\Omega\,\xi_c)^2,\nonumber\\
&&\omega=\sqrt{4\,(\Omega\,E+m\,\omega_c)^2+\eta^2_{L}},\nonumber\\
&&j=\sqrt{\eta^2_{c}-e^2\,\xi^2_{c}},\nonumber\\
&&a=2\,(e\,\xi_c\,E+m\,\eta_c),\nonumber\\
&&b=2\,[m\,\eta_L+\omega\,(l-2\,e\,\Omega\,\xi_c)],\nonumber\\
&&\omega_c=\frac{e\,B_0}{2\,m}
\label{11}
\end{eqnarray}
is called the cyclotron frequency of the particle moving in the magnetic field.

Let us define a new variable $r=\sqrt{\omega}\,x$, Eq. (\ref{10}) becomes
\begin{equation}
\left [\frac{d^2}{dr^2} + \beta-r^2-\frac{j^2}{r^2}-\frac{\eta}{r}-\theta\,r \right]\,\psi(r)=0,
\label{12}
\end{equation}
where 
\begin{equation}
\beta=\frac{\lambda}{\omega}\quad,\quad \eta=\frac{a}{\sqrt{\omega}}\quad,\quad \theta=\frac{b}{\omega^{\frac{3}{2}}}.
\label{13}
\end{equation}

We now use the appropriate boundary conditions to investigate the bound states solution in this problem. It is known in relativistic quantum mechanics that the radial wave-functions must be regular both at $r \rightarrow 0$ and $r \rightarrow \infty $. Then we proceed with the analysis of the asymptotic behavior of the radial eigenfunctions at origin and in the infinite. These conditions are necessary since the wave-functions must be well-behaved in these limit, and thus, the bound states of energy eigenvalues for this system can be obtained. Suppose the possible solution to the Eq. (\ref{12}) is
\begin{equation}
\psi (r)=r^{j}\,e^{-\frac{1}{2}\,(\theta+r)\,r}\,H(r).
\label{14}
\end{equation}
Substituting the solution Eq. (\ref{14}) into the Eq. (\ref{12}), we obtain
\begin{equation}
\frac{d^2 H}{dr^2}+\left [\frac{\gamma}{r}-\theta-2\,r \right]\,\frac{d H}{dr}+\left [-\frac{\chi}{r}+\Theta \right]\,H (r)=0,
\label{17}
\end{equation}
where
\begin{eqnarray}
&&\gamma=1+2\,j,\nonumber\\
&&\Theta=\beta+\frac{\theta^2}{4}-2\,(1+j),\nonumber\\
&&\chi=\eta+\frac{\theta}{2}\,(1+2\,j).
\label{18}
\end{eqnarray}
Equation (\ref{17}) is the biconfluent Heun's differential equation \cite{ERFM,KB,KB2,RLLV6,RLLV2,RLLV3,RLLV4,RLLV5, AR,SYS} with $H(r)$ is the Heun polynomials function.

Writing the function $H (r)$ as a power series expansion around the origin \cite{GBA}:
\begin{equation}
H(r)=\sum^{\infty}_{i=0} c_{i}\,r^{i}.
\label{19}
\end{equation}
Substituting the series solution into the Eq. (\ref{17}), we obtain the following recurrence relation:
\begin{equation}
c_{n+2}=\frac{1}{(n+2)(n+1+\gamma)}\,[\{\chi+\theta\,(n+1)\}\,c_{n+1}-(\Theta-2\,n)\,c_n].
\label{20}
\end{equation}
Few coefficients of the series solution are
\begin{eqnarray}
&&c_{1}=\left(\frac{\eta}{\gamma}+\frac{\theta}{2} \right)\,c_0,\nonumber\\
&&c_{2}=\frac{1}{2\,(1+\gamma)}\,[(\chi+\theta)\,c_1-\Theta\,c_0].
\label{21}
\end{eqnarray}

As the function $H (r)$ has a power series expansion around the origin in Eq. (\ref{19}), then, the relativistic bound states solution can be achieved by imposing that the power series expansion becomes a polynomial of degree $n$ and we obtain a finite degree polynomial for the biconfluent Heun series. Furthermore, the wave-function $\psi$ must vanish at $r \rightarrow \infty$ for this finite degree polynomial of power series otherwise the function diverge for large values of $r$. Therefore, we must truncate the power series expansion $H (r)$ a polynomial of degree $n$ by imposing the following two conditions \cite{ZW,RR1,ERFM,KB,EPJC3,KB2,RLLV6,RLLV2,RLLV3,RLLV4,RLLV5,AV,JM}:
\begin{eqnarray}
\Theta&=&2\,n,\quad (n=1,2,...)\nonumber\\
c_{n+1}&=&0. 
\label{22}
\end{eqnarray}
By analyzing the condition $\Theta=2\,n$, we have the second degree eigenvalues equation
\begin{eqnarray}
&&E^{2}_{n,l}=m^2+k^2+2\,\eta_c\,\eta_L+2\,\xi\,\Omega^2+2\,\omega\,(n+1+\sqrt{\eta^2_{c}-e^2\,\xi^2_{c}})\nonumber\\
&&-\frac{m^2\,\eta^2_{L}}{\omega^2}-\frac{2\,m\,\eta_L\,(l-2\,e\,\xi_c\,\Omega)}{\omega}.
\label{23}
\end{eqnarray}
The corresponding eigenfunctions is given by
\begin{equation}
\psi_{n,l} (r)=r^{\sqrt{\eta^2_{c}-e^2\,\xi^2_{c}}}\,e^{-\frac{1}{2}\,(r+\theta)\,r}\,H (r).
\label{24}
\end{equation}

Note that Eq. (\ref{23}) does not represent the general expression of the eigenvalue problem. One can obtain the individual energy eigenvalues one by one, that is, $E_1$, $E_2$, $E_3$,.. by imposing the additional recurrence condition
$c_{n+1}=0$ on the eigenvalues. The solution with Heun's equation makes it possible to obtain the individual eigenvalues one by one as done in \cite{ZW,RR1,ERFM,KB,EPJC3,KB2,RLLV6,RLLV2,RLLV3,RLLV4,RLLV5,AV,JM}. In order to analyze the above conditions, we must assign values to $n$. In this case, consider $n=1$, that means we want to construct a first degree polynomial to $H (r)$. With $n=1$, we have $\Theta=2$ and $c_2=0$ which implies from Eq. (\ref{21}) 
\begin{eqnarray}
&&(\chi+\theta)\,c_1=2\,c_0\Rightarrow \frac{\eta}{\gamma}+\frac{\theta}{2}=\frac{2}{\chi+\theta},\nonumber\\
&&(\frac{a_{1,l}}{1+2\,j}+\frac{b_{1,l}}{2\,\omega_{1,l}})(a_{1,l}+\frac{b_{1,l}}{\omega_{1,l}}(j+\frac{3}{2}))=2\,\omega_{1,l}\nonumber\\
&&\omega^{3}_{1,l}-\frac{a^2}{2\,(1+2\,j)}\,\omega^2_{1,l}-a\,b\,(\frac{1+j}{1+2\,j})\,\omega_{1,l}-\frac{b^2}{8}\,(3+2\,j)=0,
\label{25}
\end{eqnarray}
a constraint on the physical parameter $\omega_{1,l}$. The relation given in Eq. (\ref{25}) gives the possible values of the parameter $\omega_{1,l}$ that permit us to construct first degree polynomial to $ H (r)$ for $n=1$ \cite{ERFM,KB, KB2,RLLV6}. Note that its values changes for each quantum number $n$ and $l$, so we have labeled $\omega \rightarrow \omega_{n,l}$. In this way, we obtain the following energy eigenvalue $E_{1,l}$:
\begin{eqnarray}
E_{1,l}&=&\pm\,\{m^2+k^2+2\,\eta_c\,\eta_L+2\,\xi\,\Omega^2+2\,(2+\sqrt{\eta^2_{c}-e^2\,\xi^2_{c}})\,\omega_{1,l}\nonumber\\
&&-\frac{m^2\,\eta^2_{L}}{\omega^2_{1,l}}-\frac{2\,m\,\eta_L\,(l-2\,e\,\xi_c\,\Omega)}{\omega_{1,l}}\}^{\frac{1}{2}}.
\label{26}
\end{eqnarray}
Then, by substituting the real solution $\omega_{1,l}$ from Eq. (\ref{25}) into the Eq. (\ref{26}) it is possible to obtain the allowed values of the relativistic energy levels for the radial mode $n=1$ of a position- dependent mass system. We can see that the lowest energy state is defined by the real solution of the algebraic
equation Eq. (\ref{25}) plus the expression given in Eq. (\ref{26}) for the radial mode $n=1$, instead of $n=0$. This effect arises due to the presence of Cornell-type potential in the system. Note that, it is necessary physically that the lowest energy state is $n=1$ and not $n=0$, otherwise the opposite would imply that $c_1=0$ which is not possible.

The corresponding radial wave-function for $n=1$ is given by
\begin{equation}
\psi_{1,l} (r)=r^{\sqrt{\eta^2_{c}-e^2\,\xi^2_{c}}}\,e^{-\frac{1}{2}\,(r+\frac{b}{\omega^{\frac{3}{2}}_{1,l}})\,r}\,(c_0+c_1\,r),
\label{27}
\end{equation}
where
\begin{equation}
c_1=\frac{1}{\sqrt{\omega_{1,l}}}\,[\frac{a}{1+2\,\sqrt{\eta^2_{c}-e^2\,\xi^2_{c}}}+\frac{b}{2\,\omega_{1,l}}]\,c_0.
\label{28}
\end{equation}

\subsection{Interaction without potential}

Here we study a spin-$0$ massive charged particle by solving the Klein-Gordon equation in the presence of an external fields in a G\"{o}del-type space-time without potential and obtain the relativistic energy eigenvalue.

We choose here zero scalar and vector potentials, $S=0=V$. In that case, Eq (\ref{8}) becomes
\begin{equation}
\left [\frac{d^2}{dx^2} + E^2-m^2-k^2-l^2-2\,\xi\,\Omega^2-\omega^2\,x^2-2\,\omega\,l\,x \right]\,\psi(x)=0,
\label{bb1}
\end{equation}
The above equation can be expressed as
\begin{equation}
\left [\frac{d^2}{dx^2} + E^2-m^2-k^2-2\,\xi\,\Omega^2-\omega^2\,(x+\frac{l}{\omega})^2 \right]\,\psi(x)=0,
\label{bb2}
\end{equation}
Let us define a new variable $r=(x+\frac{l}{\omega})$, Eq. (\ref{bb2}) becomes
\begin{equation}
\psi''(r)+(\delta-\omega^2\,r^2)\,\psi(r)=0,
\label{bb3}
\end{equation}
where
\begin{equation}
\delta=E^2-m^2-k^2-2\,\xi\,\Omega^2.
\label{bb4}
\end{equation}
Again introducing a new variable $\rho=\sqrt{\omega}\,r$ into the Eq. (\ref{bb3}), we obtain
\begin{equation}
\psi''(\rho)+(\frac{\delta}{\omega}-\rho^2)\,\psi (\rho)=0
\label{bb5}
\end{equation}
which is similar a harmonic-type oscillator equation. Therefore, the energy eigenvalues equation is
\begin{eqnarray}
&&\frac{\delta}{\omega}=2\,n+1\Rightarrow \delta=(2\,n+1)\,\omega\nonumber\\\Rightarrow
&&E^2_{n}-2\,\Omega\,(2\,n+1)\,E_{n}-m^2-k^2-2\,\xi\,\Omega^2-2\,m\,\omega_c\,(2\,n+1)=0.\quad
\label{bb6}
\end{eqnarray}
The energy eigenvalues associated with $n^{th}$ modes is
\begin{eqnarray}
E_{n}&=&(2\,n+1)\,\Omega+\sqrt{(2\,n+1)^2\,\Omega^2+m^2+k^2+2\,\xi\,\Omega^2+2\,m\,\omega_c\,(2\,n+1)}\nonumber\\
&=&(2\,n+1)\,\Omega+\sqrt{(2\,n+1)^2\,\Omega^2+m^2+k^2+2\,\xi\,\Omega^2+|e\,B_0|\,(2\,n+1)},
\label{bb7}
\end{eqnarray}
where $n=0,1,2,.$. We can see that the energy eigenvalues (\ref{bb7}) depend on the parameter $\Omega$ characterizing the vorticity parameter of the space-time geometry, and the external magnetic field $B_0$ as well the non-minimal coupling constant $\xi$ with the background curvature. 

In absence of external magnetic fields, $B_0 \rightarrow 0$, and without non-minimal coupling constant, $\xi \rightarrow 0$, the eigenvalue (\ref{bb7}) becomes
\begin{equation}
E_{n}=(2\,n+1)\,\Omega+\sqrt{(2\,n+1)^2\,\Omega^2+m^2+k^2}.
\label{bb8}
\end{equation}
Equation (\ref{bb8}) is the energy eigenvalues of spin-$0$ particle in the background of a flat class of G\"{o}del-type space-time and consistent with the result in \cite{EPJC2}. Thus we can see that the energy eigenvalues (\ref{bb7}) in comparison to the result in \cite{EPJC2} get modify due to the presence of external fields and the non-minimal coupling constant with the background curvature.

Therefore, the individual energy levels for $n=0,1$ using (\ref{bb7}) are follows:
\begin{eqnarray}
&&n=0\quad:\quad E_0=\Omega+\sqrt{\Omega^2+2\,\xi\,\Omega^2+2\,m\,\omega_c+m^2+k^2},\nonumber\\
&&n=1\quad:\quad E_1=3\,\Omega+\sqrt{9\,\Omega^2+2\,\xi\,\Omega^2+6\,m\,\omega_c+m^2+k^2}.
\label{bb9}
\end{eqnarray}

A special case corresponds to $m=0=k$, the energy eigenvalues (\ref{bb7}) reduces to
\begin{equation}
E_{n}=(2\,n+1)\,\Omega+\sqrt{(2\,n+1)^2\,\Omega^2+2\,\xi\,\Omega^2+|e\,B_0|\,(2\,n+1)}.
\label{bb10}
\end{equation}
The individual energy levels for $n=0,1$ in that case are follows:
\begin{eqnarray}
&&n=0\quad:\quad E_0=\Omega+\sqrt{\Omega^2+2\,\xi\,\Omega^2+|e\,B_0|},\nonumber\\
&&n=1\quad:\quad E_1=3\,\Omega+\sqrt{9\,\Omega^2+2\,\xi\,\Omega^2+3\,|e\,B_0|}.
\label{bb11}
\end{eqnarray}
And others are in the same way. We can see that the presence of external magnetic field $B_0$ as well as the non-minimal coupling constant $\xi$ causes asymmetry in the energy levels and hence, the energy levels are not equally spaced.

The eigenfunctions is given by 
\begin{equation}
\psi_{n} (\rho)=|N|\,H_{n} (\rho)\,e^{-\frac{\rho^2}{2}},
\label{bb12}
\end{equation}
where $|N|=\sqrt{\frac{1}{2^{n}\,{n!}\,\sqrt{\pi}}}$ is the normalization constant and $H_{n} (\rho)$ are the Hermite polynomials and define as
\begin{equation}
H_{n} (\rho)=(-1)^{n}\,e^{\rho^2}\,\frac{d^{n}}{d{\rho^n}}\,(e^{-\rho^2}),\quad \int^{\infty}_{-\infty} e^{-\rho^2}\,H_{n} (\rho)\,H_{m} (\rho)\,d\rho=\sqrt{\pi}\,2^{n}\,{n!}\,\delta_{nm}.
\label{bb13}
\end{equation}

\section{\bf The Klein-Gordon Oscillator}

Here we study a spin-$0$ massive charged particle by solving the Klein-Gordon equation of the Klein-Gordon oscillator in the presence of an external fields in a G\"{o}del-type space-time subject to a Cornell-type scalar and Coulomb-type vector potentials. We analyze the effects on the relativistic energy eigenvalue and corresponding eigenfunctions due to various physical parameters.

To couple Klein-Gordon field with oscillator \cite{Bru,Dvo}, the generalization of Mirza {\it et al.} prescription \cite{Mirza}, in which the following change in the momentum operator is taken:
\begin{equation}
p_{\mu}\rightarrow p_{\mu}+i\,m\,\omega_0\,X_{\mu},
\label{cc1}
\end{equation}
where $m$ is the particle mass at rest, $\omega_0$ is the frequency of the oscillator and $X_{\mu}=(0,x,0,0)$, with $x$ being the distance of the particle. In this way, the Klein-Gordon oscillator equation becomes
\begin{equation}
\frac{1}{\sqrt{-g}}\,(D_{\mu}+m\,\omega_0\,X_{\mu})\sqrt{-g}\,g^{\mu\nu}\,(D_{\nu}-m\,\omega_0\,X_{\nu})\,\Psi=(m+S)^2\,\Psi.
\label{cc2}
\end{equation}
Using the space-time (\ref{1}), we obtain the following equation
\begin{eqnarray}
&&[-(\frac{\partial}{\partial t}+i\,e\,V)^2+\{(\frac{\partial}{\partial y}-i\,e\,A_y)-\alpha\,x\,(\frac{\partial}{\partial t}+i\,e\,V)\}^2+\frac{\partial}{\partial z^2}\nonumber\\
&&+(\frac{\partial}{\partial x}+m\,\omega_0\,x)\,(\frac{\partial}{\partial x}-m\,\omega_0\,x)-(m+S)^2]\,\Psi=0.
\label{cc3}
\end{eqnarray}
Using the ansatz (\ref{7}) into the above Eq. (\ref{cc3}), we arrive at the following equation
\begin{eqnarray}
&&\frac{d^2\,\psi}{dx^2}+[(E-e\,V)^2-\{ \alpha_0\,x\,(E-e\,V)+(l-e\,A_y) \}^2-k^2-m\,\omega_0\nonumber\\
&&-m^2\,\omega^2_{0}\,x^2-(m+S)^2]\,\psi=0.
\label{cc4}
\end{eqnarray}

Substituting the potentials Eq. (\ref{2}), (\ref{9}) and (\ref{vector}) into the equation (\ref{cc4}), we obtain the following equation:
\begin{equation}
\psi''(x)+[\tilde{\lambda}-\tilde{\omega}^2\,x^2-\frac{j^2}{x^2}-\frac{a}{x}-\tilde{b}\,x]\,\psi(x)=0,
\label{dd1}
\end{equation}
where we have defined
\begin{eqnarray}
&&\tilde{\lambda}=E^2-m^2-k^2-2\,\eta_c\,\eta_L-2\,\xi\,\Omega^2-(l-2\,e\,\Omega\,\xi_c)^2-m\,\omega_0,\nonumber\\
&&\tilde{\omega}=\sqrt{4\,(\Omega\,E+m\,\omega_c)^2+m^2\,\omega^2_{0}+\eta^2_{L}},\nonumber\\
&&j=\sqrt{\eta^2_{c}-e^2\,\xi^2_{c}},\nonumber\\
&&a=2\,(e\,\xi_c\,E+m\,\eta_c),\nonumber\\
&&\tilde{b}=2\,[m\,\eta_L+\tilde{\omega}\,(l-2\,e\,\Omega\,\xi_c)].
\label{dd2}
\end{eqnarray}
 
Let us define a new variable $r=\sqrt{\tilde{\omega}}\,x$, Eq. (\ref{dd1}) becomes
\begin{equation}
\left [\frac{d^2}{dr^2} +\tilde{\beta}-r^2-\frac{j^2}{r^2}-\frac{\tilde{\eta}}{r}-\tilde{\theta}\,r \right]\,\psi(r)=0,
\label{dd7}
\end{equation}
where 
\begin{equation}
\tilde{\beta}=\frac{\lambda}{\tilde{\omega}}\quad,\quad \tilde{\eta}=\frac{a}{\sqrt{\tilde{\omega}}}\quad,\quad \tilde{\theta}=\frac{\tilde{b}}{\tilde{\omega}^{\frac{3}{2}}}.
\label{dd8}
\end{equation}

Suppose the possible solution to the Eq. (\ref{dd7}) is
\begin{equation}
\psi (r)=r^{j}\,e^{-\frac{1}{2}\,(\tilde{\theta}+r)\,r}\,H(r).
\label{dd9}
\end{equation}
Substituting the solution Eq. (\ref{dd9}) into the Eq. (\ref{dd7}), we obtain
\begin{equation}
H'' (r)+[\frac{\gamma}{r}-\tilde{\theta}-2\,r]\,H' (r)+[-\frac{\tilde{\chi}}{r}+\tilde{\Theta}]\,H (r)=0,
\label{dd3}
\end{equation}
where $\gamma$ is given earlier and
\begin{eqnarray}
&&\tilde{\Theta}=\tilde{\beta}+\frac{\tilde{\theta}^2}{4}-2\,(1+j),\nonumber\\
&&\tilde{\chi}=\tilde{\eta}+\frac{\tilde{\theta}}{2}\,(1+2\,j)
\label{dd4}
\end{eqnarray}
Equation (\ref{dd3}) is the biconfluent Heun's differential equation \cite{ERFM,KB,KB2,RLLV6,RLLV2,RLLV3,RLLV4,RLLV5, AR,SYS}.

Substituting the series solution Eq. (\ref{19}) into the Eq. (\ref{dd3}), we obtain the following recurrence relation:
\begin{equation}
c_{n+2}=\frac{1}{(n+2)(n+1+\gamma)}\,[\{\tilde{\chi}+\tilde{\theta}\,(n+1)\}\,c_{n+1}-(\tilde{\Theta}-2\,n)\,c_n].
\label{dd10}
\end{equation}
Few coefficients of the series solution are
\begin{eqnarray}
&&c_{1}=\left(\frac{\tilde{\eta}}{\gamma}+\frac{\tilde{\theta}}{2} \right)\,c_0,\nonumber\\
&&c_{2}=\frac{1}{2\,(1+\gamma)}\,[(\tilde{\chi}+\tilde{\theta})\,c_1-\tilde{\Theta}\,c_0].
\label{dd11}
\end{eqnarray}
The power series expansion becomes a polynomial of degree $n$ by imposing following two conditions \cite{ZW,RR1,ERFM,KB,EPJC3,KB2,RLLV6,RLLV2,RLLV3,RLLV4,RLLV5,AV,JM}:
\begin{eqnarray}
\tilde{\Theta}&=&2\,n \quad  (n=1,2,..)\nonumber\\
c_{n+1}&=&0
\label{dd12}
\end{eqnarray}

Using the first condition we obtain the following energy eigenvalues :
\begin{eqnarray}
&&E^{2}_{n,l}=m^2+k^2+2\,\eta_c\,\eta_L+2\,\xi\,\Omega^2+m\,\omega_0+2\,(n+1+\sqrt{\eta^2_{c}-e^2\,\xi^2_{c}})\,\tilde{\omega}\nonumber\\
&&-\frac{m^2\,\eta^2_{L}}{\tilde{\omega}^2}-\frac{2\,m\,\eta_L\,(l-2\,e\,\xi_c\,\Omega)}{\tilde{\omega}}.
\label{dd5}
\end{eqnarray}
The corresponding eigenfunctions is given by
\begin{equation}
\psi_{n,l}(r)=r^{\sqrt{\eta^2_{c}-e^2\,\xi^2_{c}}}\,e^{-\frac{1}{2}\,(r+\tilde{\theta})\,r}\,H (r).
\label{dd6}
\end{equation}

As done earlier, we obtain the individual energy levels by imposing the recurrence condition $c_{n+1}=0$. For $n=1$, we have $c_2=0$ which implies from Eq. (\ref{dd11})
\begin{eqnarray}
\tilde{\omega}^{3}_{1,l}-\frac{a^2}{2\,(1+2\,j)}\,\tilde{\omega}^2_{1,l}-a\,\tilde{b}\,(\frac{1+j}{1+2\,j})\,\tilde{\omega}_{1,l}-\frac{\tilde{b}^2}{8}\,(3+2\,j)=0,
\label{dd13}
\end{eqnarray}
a constraint on the physical parameter $\tilde{\omega}_{1,l}$. The relation given in Eq. (\ref{dd13}) gives the possible values of the parameter $\tilde{\omega}_{1,l}$ that permit us to construct first degree polynomial to $ H (r)$ for $n=1$ \cite{ERFM,RLLV6,KB,KB2}. In this way, we obtain the following second degree algebraic equation for $E_{1,l}$:
\begin{eqnarray}
&&E_{1,l}=\pm\,\{m^2+k^2+2\,\eta_{c}\,\eta_{L}+2\,\xi\,\Omega^2+2\,(2+\sqrt{\eta^2_{c}-e^2\,\xi^2_{c}})\,\tilde{\omega}_{1,l}\nonumber\\
&&-\frac{m^2\,\eta^2_{L}}{\tilde{\omega}^2_{1,l}}-\frac{2\,m\,\eta_{L}\,(l-2\,e\,\xi_{c}\,\Omega)}{\tilde{\omega}_{1,l}}\}^{\frac{1}{2}}.
\label{dd14}
\end{eqnarray}
Then, by substituting the real solution $\tilde{\omega}_{1,l}$ from Eq. (\ref{dd13}) into the Eq. (\ref{dd14}) it is possible to obtain the allowed values of the relativistic energy levels for the radial mode $n=1$ of a position- dependent mass system. We can see that the lowest energy state is defined by the real solution of algebraic equation Eq. (\ref{dd13}) plus the expression given in Eq. (\ref{dd14}) for the radial mode $n=1$, instead of $n=0$. This effect arises due to the presence of Cornell-type potential in the system. Note that, it is necessary physically that the lowest energy state is $n=1$ and not $n=0$, otherwise the opposite would imply that $c_1=0$ which is not possible.

The corresponding radial wave-function for $n=1$ is given by
\begin{equation}
\psi_{1,l} (r)=r^{\sqrt{\eta^2_{c}-e^2\,\xi^2_{c}}}\,e^{-\frac{1}{2}\,(r+\frac{b}{\tilde{\omega}^{\frac{3}{2}}_{1,l}})\,r}\,(c_0+c_1\,r),
\label{dd15}
\end{equation}
where
\begin{equation}
c_1=\frac{1}{\sqrt{\tilde{\omega}_{1,l}}}\,[\frac{2\,(e\,\xi_c\,E_{1,l}+m\,\eta_c)}{1+2\,\sqrt{\eta^2_{c}-e^2\,\xi^2_{c}}}+\frac{2\,\left(m\,\eta_L+\omega_{1,l}\,(l-2\,e\,\Omega\,\xi_c) \right)}{2\,\tilde{\omega}_{1,l}}]\,c_0.
\label{dd16}
\end{equation}

\section{Conclusions}

The relativistic quantum system of scalar and spin-half particles in G\"{o}del-type space-times was investigated by several authors ({\it e. g.}, \cite{ND,SD,JC,ZW,EPJC2,MPLA,RR1,EPJC3,RLLV6}). They demonstrated that the energy eigenvalues of the relativistic quantum system get modified and depend on the global parameters characterizing the space-times.

In this work, we have investigated the influence of vorticity parameter on the relativistic energy eigenvalues of a relativistic scalar particle in a G\"{o}del-type space-time subject to a homogeneous magnetic field with potentials. We have derived the radial wave-equation of the Klein-Gordon equation in a class of flat G\"{o}del-type space-time in the presence of an external fields with(-out) potentials by choosing a suitable ansatz of the wave-function. In {\it sub-section 2.1}, we have introduced a Cornell-type scalar and Coulomb-type vector potentials into the considered relativistic system and obtained the energy eigenvalue Eq. (\ref{23}) and corresponding eigenfunctions Eq. (\ref{24}). We have seen that the presence of a uniform magnetic field and potential parameters modifies the energy spectrum in comparison those result obtained in \cite{EPJC2}. By imposing the additional recurrence condition $c_{n+1}=0$, we have obtained the ground state energy levels Eq. (\ref{26}) and wave-functions Eq. (\ref{27})--(\ref{28}) for $n=1$. In {\it sub-section 2.2}, we have considered zero potential into the relativistic system and solved the radial wave-equation of the Klein-Gordon equation in the presence of an external field. We obtained the energy eigenvalues Eq. (\ref{bb7}) and compared with the results obtained in \cite{EPJC2}. We have seen that the relativistic energy eigenvalues Eq. (\ref{bb7}) get modify in comparison to those in \cite{EPJC2} due to the presence of a homogeneous magnetic field. In {\it section 3}, we have solved the Klein-Gordon equation of the Klein-Gordon oscillator in a G\"{o}del-type space-time subjected to a homogeneous magnetic field in the presence of a Cornell-type scalar and Coulomb-type vector potentials. We have obtained the energy eigenvalue Eq. (\ref{dd5}) and corresponding eigenfunctions Eq. (\ref{dd6}). We have seen that the presence of a uniform magnetic field and potential parameters modifies the energy spectrum in comparison to those in \cite{EPJC2}. By imposing the additional recurrence condition $c_{n+1}=0$, we have obtained the ground state energy levels Eq. (\ref{dd14}) and wave-function Eq. (\ref{dd15}) for $n=1$ and others are in the same way.

So, in this paper we have some results which are in addition to the previous results obtained in \cite{ND,SD,JC,ZW,EPJC2,MPLA,EPJC3,CTP2,RLLV6} present may interesting effects. This is the fundamental subject in physics and the connection between these theories (quantum mechanics and gravitation) are not well understood.

\section*{Data Availability}

No data has been used to prepare this paper.

\section*{Conflict of Interest}

Author declares that there is no conflict of interest regarding publication this paper.

\section*{Acknowledgment}

Author sincerely acknowledge the anonymous kind referee(s) for his/her valuable comments and suggestions.

\end{document}